\documentclass{aastex631}

\usepackage{color,xcolor}
\usepackage{float} 
\usepackage{comment}
\usepackage{threeparttable}
\usepackage{graphicx} 
\usepackage{txfonts}
\usepackage{hyperref}

\hypersetup{colorlinks,linkcolor={blue},citecolor={blue},urlcolor={blue}} 

\begin{document}

\title{Radio-loudness Statistics of Quasars from Quaia--VLASS}

\author[0009-0009-8592-5400]{Nestor Arsenov}
\email{narsenov@nao-rozhen.org}

\affiliation{Institute of Astronomy and NAO, Bulgarian Academy of Sciences \\
72 Tsarigradsko Chaussee Blvd. \\
1784 Sofia, Bulgaria}
\affiliation{MTA--CSFK \emph{Lend\"ulet} ``Momentum'' Large-Scale Structure (LSS) Research Group\\
Konkoly Thege Mikl\'os \'ut 15-17\\
H-1121 Budapest, Hungary}

\author[0000-0003-3079-1889]{S\'{a}ndor Frey}
\affiliation{Konkoly Observatory, HUN-REN Research Centre for Astronomy and Earth Sciences\\
Konkoly Thege Mikl\'os {\'u}t 15-17\\
H-1121 Budapest, Hungary}
\affiliation{CSFK, MTA Centre of Excellence\\
Konkoly Thege Mikl\'os {\'u}t 15-17\\
H-1121 Budapest, Hungary}
\affiliation{Institute of Physics and Astronomy, ELTE E\"otv\"os Lor\'and University\\
P\'azm\'any P\'eter s\'et\'any 1/A\\
H-1117 Budapest, Hungary}

\author[0000-0002-5825-579X]{Andr\'as Kov\'acs}
\affiliation{MTA--CSFK \emph{Lend\"ulet} ``Momentum'' Large-Scale Structure (LSS) Research Group\\
Konkoly Thege Mikl\'os \'ut 15-17\\
H-1121 Budapest, Hungary}
\affiliation{Konkoly Observatory, HUN-REN Research Centre for Astronomy and Earth Sciences\\
Konkoly Thege Mikl\'os {\'u}t 15-17\\
H-1121 Budapest, Hungary}
\affiliation{CSFK, MTA Centre of Excellence\\
Konkoly Thege Mikl\'os {\'u}t 15-17\\
H-1121 Budapest, Hungary}

\author[0000-0002-1582-4913]{Lyuba Slavcheva-Mihova}
\affiliation{Institute of Astronomy and NAO, Bulgarian Academy of Sciences \\
72 Tsarigradsko Chaussee Blvd. \\
1784 Sofia, Bulgaria}

\begin{abstract}

Quasars are objects of high interest in extragalactic astrophysics, cosmology, and astrometry. One of their useful qualities is their potential radio loudness. However, the fraction of radio-loud vs. radio-quiet quasars is subject to ongoing investigations, where the statistical power is limited by the low number of known quasars with radio counterparts. In this analysis, we revisited the radio loudness statistics of quasars by significantly expanding the pool of known sources. Our main goal was to create a new, value-added quasar catalog with information about their extinction-corrected magnitudes, radio flux density, possible contamination levels, and other flags, besides their sky coordinates and photometric redshifts. We cross-matched the optical Quaia catalog of about $1.3$ million quasars (from the Gaia data) with $1.9$ million sources from the Very Large Array Sky Survey (VLASS) radio catalog. We explored different thresholds for the matching radius, balancing completeness and purity of the resulting Quaia--VLASS catalog, and found $1.5\arcsec$ a sufficient choice. Our main finding is that the quasar radio-loud fraction is in good agreement with previous works ($<10\%$), and there is no significant large-scale sky pattern in radio loudness. The exact estimate depends on the $G$-band magnitude limit, and we observed weak trends with redshift and absolute optical magnitude, possibly indicating remnant systematic effects in our data. The cross-matched Quaia--VLASS catalog with 43,650 sources and accompanying code are available at \href{https://doi.org/10.5281/zenodo.16035690}{10.5281/zenodo.16035690}. This latest census of QSOs with radio counterparts will facilitate further investigations of the dichotomy of radio-loud and radio-quiet quasars, and it may also support other lines of research.

\end{abstract}


\section{Introduction}
\label{sec:intro}

Active galactic nuclei (AGN) are among the most powerful and dynamic objects in the Universe \citep[e.g.][]{2017A&ARv..25....2P,2019ApJ...871..258S,2022Natur.609..265J}, some of which can outshine their host galaxies. They can be found over a wide range of distances from the local Universe up to extremely high redshifts \citep[$z \gtrsim 10$,][]{2023ApJ...955L..24G}. AGN play an important role in astrophysics, astrometry, and cosmology. They are located at the centres of galaxies and powered by accretion onto supermassive black holes (SMBHs) which range from millions to billions of solar masses. As material spirals into the SMBH, it heats up and emits huge amounts of energy across the entire electromagnetic spectrum, from radio to $\gamma$-rays \citep[e.g.][]{2017FrASS...4...35P}.

Members of a small but spectacular subset of AGN exhibit intense radio emission originating from synchrotron radiation of relativistic charged particles moving in magnetized plasma jets launched from the close vicinity of SMBHs \citep[e.g.][]{1995PASP..107..803U,2019ARA&A..57..467B}. Studying these so-called radio-loud, or rather jetted \citep{2017NatAs...1E.194P} AGN is crucial for several reasons. Their jets are among the most extreme particle accelerators in the Universe, giving unique insight into relativistic effects, strong magnetic fields, and high-energy particle interactions. The radiative and mechanical feedback of AGN jets can influence the evolution of their host galaxies by regulating star formation \citep[e.g.][]{2012ARA&A..50..455F,2020NewAR..8801539H}. 

Owing to their presence in the widest range of cosmological distances, AGN are excellent probes of the early Universe and offer the possibility of contributing to the studies of cosmological models \citep[e.g.][]{1994ApJ...430..467V,1999A&A...342..378G}. AGN also trace cosmic structures and thus are important for understanding the large-scale distribution of matter \citep[see e.g.][]{Kovacs2021,Piccirilli2024}. In astrometry, bright radio-emitting AGN define the most precise reference frame through high-resolution very long baseline interferometry (VLBI) observations \citep{2020A&A...644A.159C}.

A subclass of AGN, quasars (originally meaning quasi-stellar radio sources) were discovered as distant extragalactic objects in the early 1960s \citep{1963Natur.197.1040S}. Not much later, it turned out that the majority of the so-called quasi-stellar objects (QSOs) are not strong radio emitters \citep{1965ApJ...141.1560S}. Nowadays, the number of known quasars with measured spectroscopic redshifts exceeds 1 million \citep{2023OJAp....6E..49F}. The fraction of radio-loud quasars among the whole population is known to be $\lesssim 10\%$ \citep[e.g.][]{2002AJ....124.2364I,2016ApJ...831..168K}.

There are various working definitions of radio loudness in the literature \citep{1989AJ.....98.1195K, 2002AJ....124.2364I, 2012ApJ...759...30B, macfarlane2021radio, yue2024novel}. One can use the radio-to-optical flux density ratios as originally proposed by \citet{1970ApJ...162..371S}. Perhaps the most widely used definition is based on the ratio of the rest-frame $5$-GHz radio and 4400-\AA\ optical flux densities \citep{1989AJ.....98.1195K}. If this ratio $R$ exceeds 10 the object is considered radio-loud. \citet{1980ApJ...238..435S} used 2500-\AA\ optical flux densities. Given the well-known flux density variability of quasars and the fact that optical and radio measurements are rarely contemporaneous, $R$ values themselves can be uncertain. 

The above two definitions are broadly consistent with each other for the quasar class as a whole \cite[see a recent example in][]{2024A&A...690A.321K}. Alternatively, dictated by practical aspects such as the availability of measurements, it is customary to use other radio and optical bands, e.g. $1.4$~GHz in the radio and the $i$-band in the optical \citep[like in][]{2002AJ....124.2364I,2012ApJ...759...30B}. Note that \citet{2003ApJ...583..145T} defined radio loudness as the ratio of radio ($5$~GHz) and X-ray ($2-10$~keV band) luminosities that can be applied for heavily obscured AGN. Similarly, \cite{klindt2019fundamental} introduced a radio-loudness metric based on the ratio of rest-frame $1.4$~GHz radio luminosity to $6~\mathrm{\mu m}$ mid-infrared luminosity, which is less affected by dust extinction than traditional optical-based definitions. A fundamentally different approach is defining radio loudness by setting a threshold based on the (monochromatic) radio luminosity \citep[e.g.][]{1986MNRAS.218..265P,1990MNRAS.244..207M}. Interestingly, the later definition and the one by \citet{2003ApJ...583..145T} provide generally similar classifications, as discussed in Appendix~C of \citet{2002AJ....124.2364I}.

\begin{figure*}
\centering
\includegraphics[width=180mm]{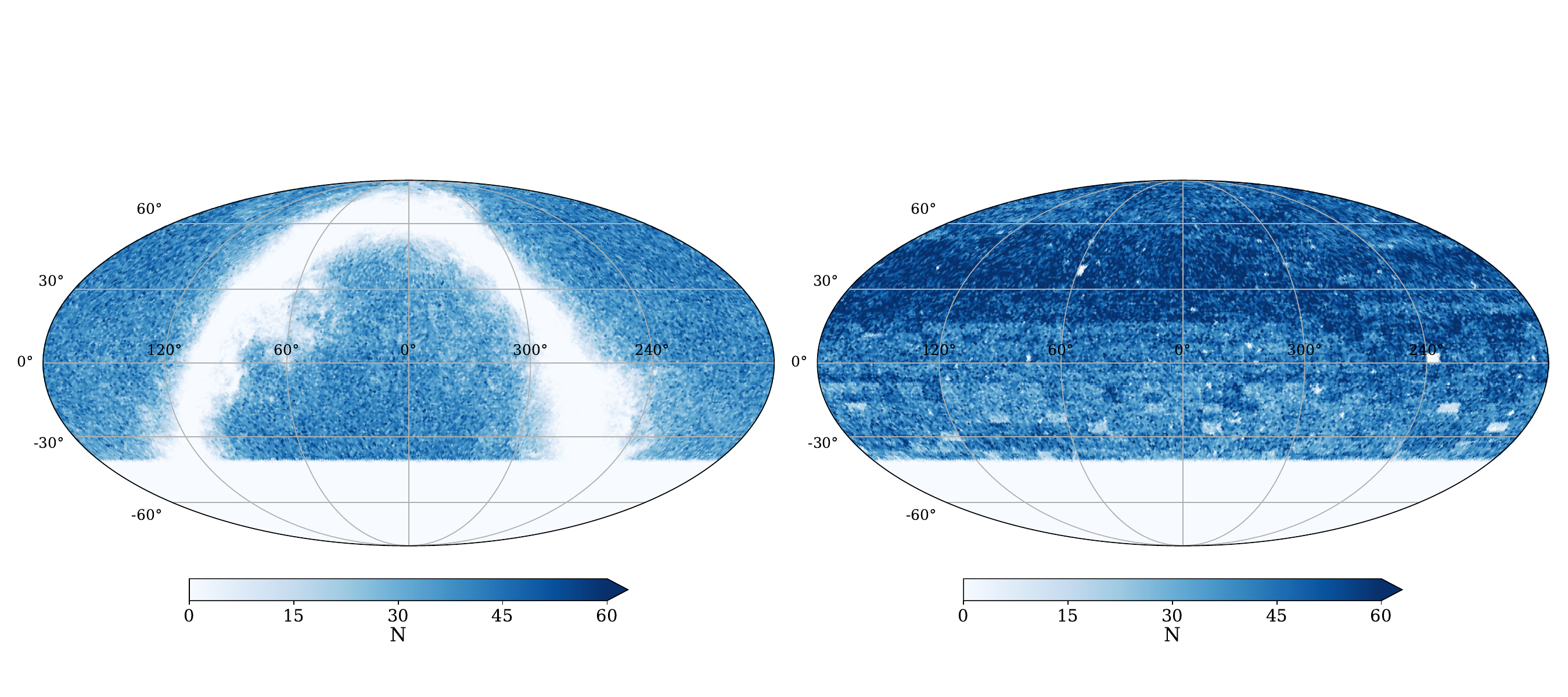}
\caption{On the left the distribution of the $1.1$ million Quaia sources that fall within the joint coverage with the VLASS survey ($\delta>-40\degr$). The dust obscuration in the Milky Way disc results in the detection of fewer QSOs in the optical. On the right the distribution of the $1.9$ million VLASS2 QL sources after applying quality flags (see Table \ref{tab:cuts_n_radio loudness}). Unlike for optical QSOs, the radio coverage is virtually unaffected by the Galactic disc.}
\label{fig:sky_map_quaia_vlass}
\end{figure*}

The `classical' definition of radio loudness based on radio and optical flux density ratios is the best applicable for quasars where the optical emission is dominantly related to the accretion disk while the radio emission comes mainly from the jets \citep{2017NatAs...1E.194P}. Nonetheless, \cite{macfarlane2021radio} show that for radio-quiet quasars the majority of the radio emission can come from star formation, and \cite{panessa2022hard} show that other forms of emission can contribute significantly to the radio flux of radio-quiet quasars, e.g. AGN winds. Most early publications found a bimodal distribution of radio-loud and radio-quiet (in fact, not totally free from radio emission, rather radio-weak) quasars \citep[e.g.][]{1980A&A....88L..12S,1989AJ.....98.1195K,1990MNRAS.244..207M,2002AJ....124.2364I}. While the debate about this dichotomy is not settled yet, a consensus seems to emerge that radio loudness has a continuous distribution with no sharp gap between radio-loud and radio-quiet sources \citep[see a review in][]{2016ApJ...831..168K, 2019A&A...622A..11G, macfarlane2021radio, 2024MNRAS.528.4547A, yue2024novel}. 
 
For a long time, studies of radio loudness were naturally limited to small samples of relatively bright AGN. The emergence of sensitive large radio and optical sky surveys \citep[e.g.][]{2002AJ....124.2364I,2019A&A...622A..11G,2024MNRAS.528.4547A} made substantially larger samples available for investigations (see Table \ref{tab:comparison_previous_works}). Here we revisit the radio loudness and fraction of radio-loud QSOs with two modern optical and radio surveys and we test statistically the dependence of radio loudness on redshift, absolute optical magnitude and selection function (later explained in Section \ref{subsec:quasar_cat}).

The paper is organized as follows. In Section \ref{sec:data}, we introduce our observational data sets and the analysis methodology. Then, Section \ref{sec:res} contains our observational results, followed by a summary of our main conclusions in Section \ref{sec:disc}.

Throughout this work, we use a $\Lambda$CDM cosmology, including our calculations of luminosity distances $d_{\rm lm}(z)$ with \texttt{astropy}\footnote{\url{https://www.astropy.org/}}, based on \citet{Planck2018} parameter values: $H_0 = 67.66\,\mathrm{km\,s^{-1}\,Mpc^{-1}}$, $\Omega_{\rm m}= 0.31$ and $\Omega_{\Lambda} = 0.69$.

\section{Data Sets and Methodology}
\label{sec:data}

The main goal of this work is to cross-match a large sample of optically detected quasars with a catalog of radio sources, mainly following the footsteps of \citet{2002AJ....124.2364I}, concentrating on the radio loudness statistics. In this Section, we present the input data sets and the technical details of the cross-matching process.

For the optical quasar catalog, we chose the recently compiled Quaia \citep{Storey_Fisher_2024} because of its large sky coverage, high completeness, and high astrometric accuracy. There are several recent radio surveys available for the given goals of this work, e.g. the Very Large Array Sky Survey \citep[VLASS,][]{2020PASP..132c5001L} and the LOFAR Two-metre Sky Survey (LoTSS) \citep{shimwell2017lofar}. The latter operates at 144 MHz and offers high sensitivity for synchrotron-dominated emission, making it particularly valuable for studying extended radio AGN and jet structures. We decided to use VLASS, which operates at a higher frequency (3 GHz) and is superior to LoTSS in positional accuracy, angular resolution and size of covered sky area. 

\subsection{Quasar Catalog}
\label{subsec:quasar_cat}

For our cross-matched sample, we used the Quaia quasar catalog \citep{Storey_Fisher_2024}, culled from the Gaia space telescope's data set \citep{Gaia2016}. The QSOs of the Quaia catalog are distinguished from the other types of sources by cross-matching the Gaia Data Release 3 (DR3) QSO candidates \citep{2023A&A...674A...1G,2023A&A...674A..41G, 2023A&A...674A..31D} with infrared sources from the Wide-field Infrared Survey Explorer \citep[WISE,][]{2014AJ....147..108L}, applying color cuts within the Gaia--WISE photometry parameter space and cutting out sources with high proper motions. The Quaia redshifts, that we use through out this work, are estimated by \cite{Storey_Fisher_2024} with a k-Nearest-Neighbors (kNN) model, using as training features colors from the Gaia and WISE photometry, the Gaia $G$-band magnitude, the dust reddening $E(B - V)$ at the location of the sources, and cruder estimates for the redshifts of the QSOs from \cite{2023A&A...674A..41G}. The labels for the kNN training are SDSS redshift values, for the subset of QSOs with available measurements.

More precisely, we used the Quaia catalog restricted in magnitude to $G<20.5$, containing about $1.3$ million QSOs, which yields high completeness of quasars within the Gaia sky coverage. There is also a conservative Quaia catalog, a subsample of the larger one, with a constraint of $G<20.0$, with the benefit of the latter constraint yielding a purer QSO sample with a higher fidelity of the sources being QSOs. We decided to use the more complete catalog with $G<20.5$, because its contaminants are expected to be mostly stars and should have negligible emission in the radio. 

Another Quaia data product that we used is the so-called selection function. It is a healpix \citep{gorski2005healpix} all-sky map with pixel values in the range $[0,1]$. The values indicate to what extent the QSO count completeness in the particular pixel has been diminished by the most severe systematics -- dust, the source density of the parent surveys, and the scan patterns of the parent surveys. Hence the selection function is a useful tool to correct for these systematics in a statistical analysis. The sky coverage of the Quaia catalog that we consider in this analysis is displayed in Figure \ref{fig:sky_map_quaia_vlass}. 

The Quaia catalog has several limitations, which have to be taken into account. On the one side, the redshift estimation for Quaia is spectrophotometric and superior to the initial Gaia redshift estimates, but inferior to spectroscopic ones, like those from SDSS \citep{Lyke_2020}. Users should consider the redshift uncertainties and the presence of catastrophic outliers in Quaia. The Quaia selection function carries limitations, too: diminished accuracy close to the Galactic disc, the inevitability to mask out Galactic regions for some tasks, dependence on right ascension and declination, but not redshift, assumption of isotropy, which is contested \citep{1967Natur.216..748S, 2021ApJ...908L..51S}. For a deeper discussion of the Quaia limitations, see \cite{Storey_Fisher_2024}. Our work is independent of some of the limitations, mostly regarding cosmological applications. For instance, the lower number density of Quaia compared to other surveys, like SDSS, is not a concern in this work, nor are isotropy assumptions or the lack of homogeneity in the distribution of Quaia or VLASS sources. The radio-loud fractions (RLFs) presented in Table \ref{tab:cuts_n_radio loudness} are not influenced by limitations in Quaia's redshift or selection function values, nor is the cross-matching of Quaia and VLASS, since it depends only on the right ascension and declination values. Nevertheless, potential Quaia miss-classifications of galaxies as QSOs would lead to impurities in our data product of Quaia--VLASS quasar cross-matches. Evidence for such impurities is discussed in Section \ref{sec:Radio_Loudness_vs_QSO_Completeness_and_Redshift}. A further limitation of Quaia is that the uncertainty in the redshifts makes the search for correlation of properties with redshift and $G_\mathrm{abs}$ more difficult, as discussed in Section \ref{sec:Radio_Loudness_vs_QSO_Completeness_and_Redshift}. Not mentioned by \cite{Storey_Fisher_2024}, the $G$-band is not extinction-corrected, which we take care of in Section \ref{subsec:reddening-correction} and observe a strong improvement of the extinction bias.

\begin{figure*}
\centering
\includegraphics[width=175mm]{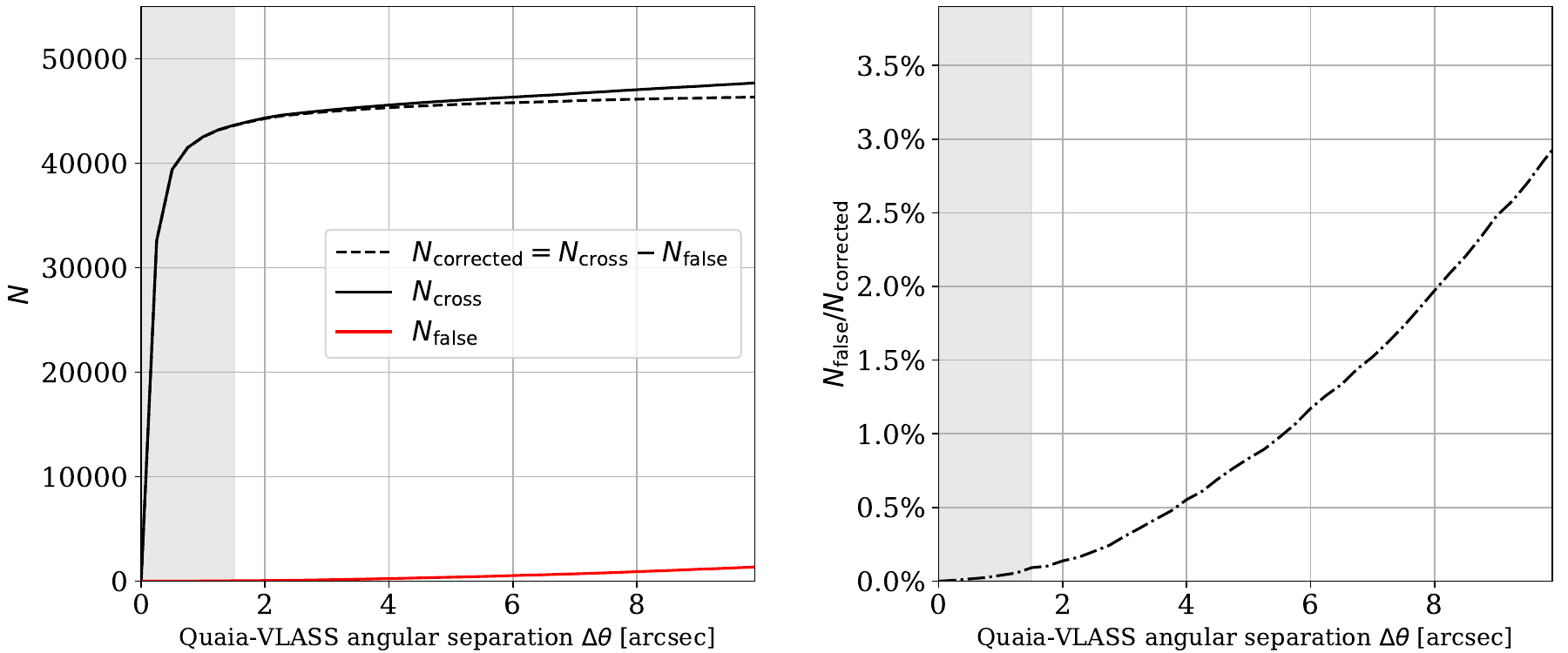}
\caption{\textit{Left}: the number of cross-matched sources as a function of a search radius parameter in arcseconds. We performed a robustness test using randomly perturbed sky coordinates, and found that a search radius of $1.5\arcsec$ results in a low level of contamination ($<0.1\%$) from false matches (\textit{right}). We thus choose this threshold for our cross-matched Quaia--VLASS sample.}
\label{fig:ang_sep_cumulative_hist}
\end{figure*}

\subsection{Radio Sources}

The other data set that we used is based on the ongoing VLASS \citep{2020PASP..132c5001L} at 3 GHz (S-band), conducted with the Very Large Array in New Mexico, in particular the VLASS Epoch 2 Quick Look (VLASS2 QL) catalog. Unlike Quaia, the VLASS2 QL catalog does not cover the full sky, and its 3.0 million radio sources are distributed at declinations $\delta>-40\degr$. 

Further, VLASS is conducted in a single band, making it more difficult to deduce the class of the detected objects than for optical surveys with multiple bands, e.g. by segmentation in color--color space. Hence, cross-matching VLASS with another large-scale catalog like Quaia is a straightforward choice for sampling radio-loud extragalactic objects. The sky area density of the VLASS2 QL catalog, with applied quality flags, is displayed on the right in Figure \ref{fig:sky_map_quaia_vlass}.

\begin{figure}
\centering
\includegraphics[width=90mm]{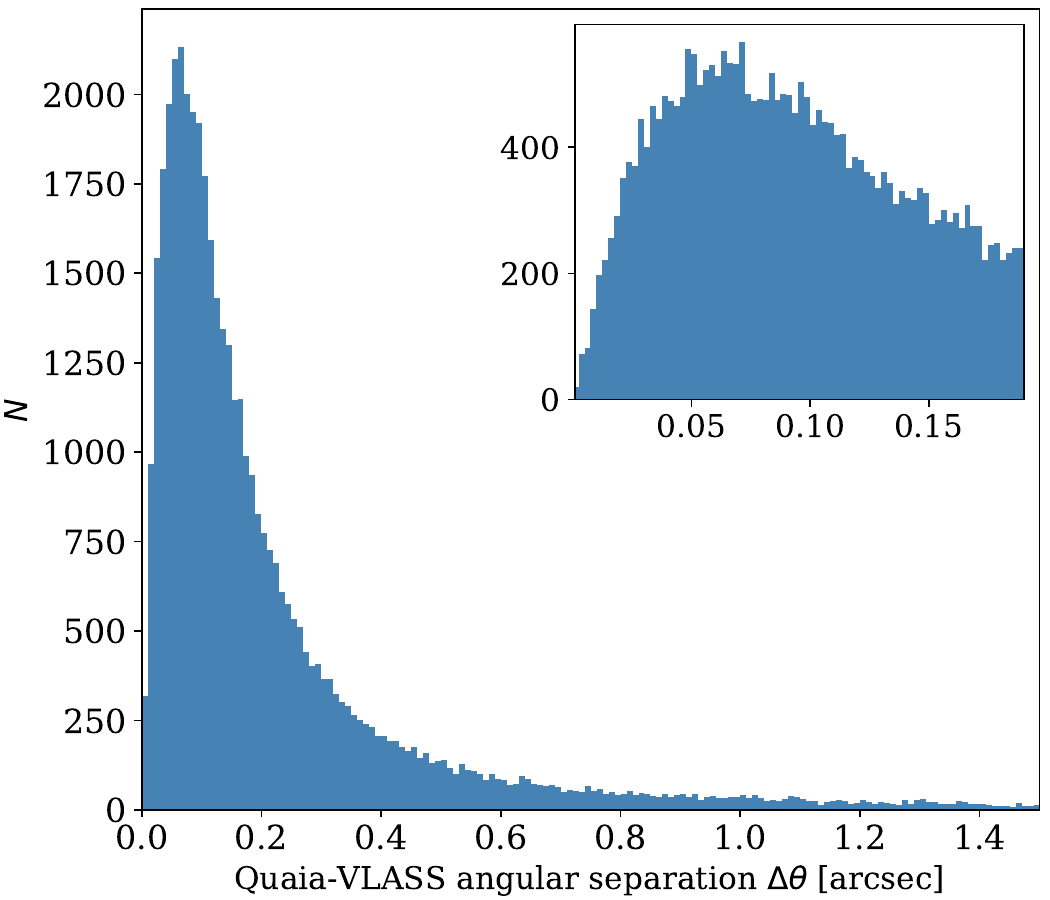}
\caption{The distribution of angular separation values when cross-matching the Quaia and VLASS samples, up to the $1.5\arcsec$ search radius threshold that we applied. The inset shows the closest matches below $\Delta \theta = 0.2\arcsec$.}
\label{fig:ang_sep_hist}
\end{figure}

For our purposes, we applied the VLASS data flags explicitly recommended by the user manual \citep{Gordon2023} to restrict ourselves to the most reliable data. Any duplicated radio sources, originating from overlaps of the sky patches observed in the survey, were excluded (\textit{Duplicate\_flag$<$2}). Further, we only used VLASS sources with no quality issues (\textit{Quality\_flag}==0) and sources with integrated flux density less than the peak brightness, but everything else in the measurement acceptable ((\textit{Quality\_flag}==4)\&(\textit{S\_Code}$\neq$‘E’)). Table~\ref{tab:cuts_n_radio loudness} shows the number of sources within the input data sets, after the quality cuts and the different $G$ magnitude bins that we used.

\begin{table}
\caption{Population statistics for the different data sets used.}
\label{tab:cuts_n_radio loudness}
\centering                          \begin{tabular}{lcccc}
\hline \hline
Source catalogs & $N_\mathrm{sources}$ & $N_\mathrm{loud}$ & $N_\mathrm{Quaia}$ & RLF ($N_\mathrm{loud}/N_\mathrm{Quaia}$) \\
\hline
\text{VLASS2 QL} & 2995025 & - & - & - \\
\hspace{1em}\text{VLASS2 QL, flags applied} & 1873524 & - & - & - \\
\text{Quaia, full ($G<20.5$)} & 1295502 & - & - & - \\
\hspace{1em}\text{Quaia, $\delta>-40^\circ$ cut} & 1084545 & - & - & - \\
\hline
\hspace{2em} \text{Quaia--VLASS matches (all)} & 43650 & 41496 & 1084545 & 3.83\% $\pm$ 0.02\% \\
\hspace{3em}14.0 $<G <$ 18.5 & 9395 & 7431 & 92397 & 8.04\% $\pm$ 0.09\% \\
\hspace{3em}18.5 $\leq G <$ 19.5 & 15763 & 15576 & 330779 & 4.71\% $\pm$ 0.04\% \\
\hspace{3em}19.5 $\leq G <$ 20.5 & 18492 & 18489 & 661369 & 2.8\% $\pm$ 0.02\% \\
\hline
\end{tabular}
\tablecomments{\textit{Top:} the number of sources contained in the data sets that we used. \textit{Bottom:} The RLF of sources in the full Quaia--VLASS cross-match and in subsets due to different cuts in $G$ optical magnitude (errors are based on Poisson statistics). An indented row is a subsample of the previous non-indented row. $N_\mathrm{loud}$ is the number of radio loud sources among the matches ($N_{sources}$) in the given row. $N_\mathrm{Quaia}$ is the number of Quaia sources used for yielding the matches.}
\end{table}

\subsection{Cross-matching}
\label{subsec:Cross-matching}

In order to create a cross-matched sample, we determined the common, physically identical sources within Quaia and VLASS. We first identified for each Quaia source its closest counterpart in VLASS and then calculated the corresponding angular separation between their sky coordinates ($\Delta \theta$). This method has been previously used, among others, by \citet{2002AJ....124.2364I,Orosz_2013}. 

\begin{figure*}
\centering
\includegraphics[width=170mm]{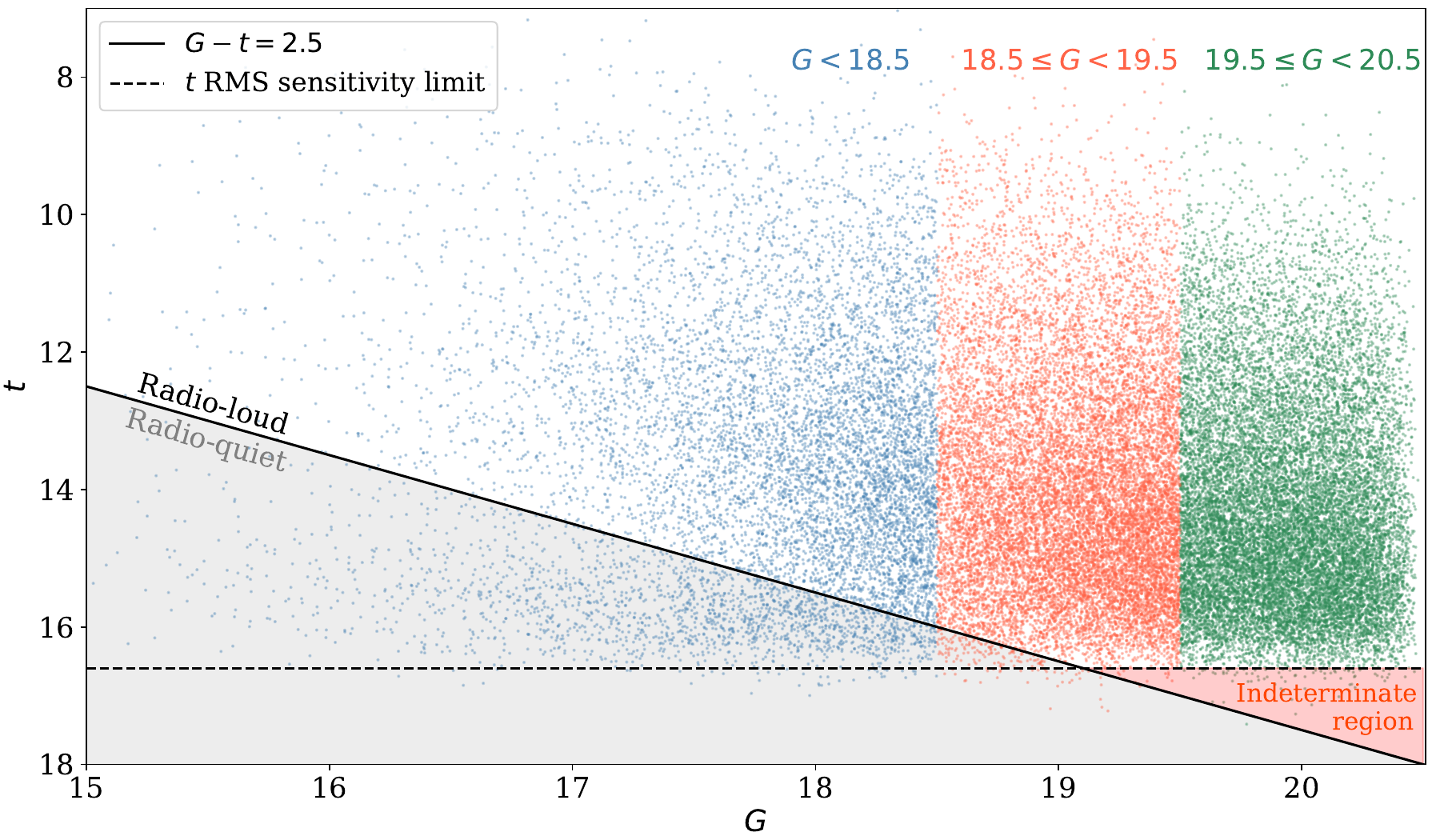}
\caption{The extent of our Quaia--VLASS sample along the $G$ (optical) and $t$ (radio) bands. Radio-loud and radio-quiet sources are separated by the black line. We defined three bins in the $G$ band, which we later analyse separately in the rest of the paper when looking for brightness-dependent trends in radio loudness. The dashed line shows the $t=16.6$ sensitivity limit, as inferred from the RMS noise limit stated in \cite{lacy2022vlass}, which agrees reasonably well with the lack of sources in the figure at $t>16.6$. The Quaia catalog has a limit at $G=20.5$ as imposed by \cite{Storey_Fisher_2024}. The region within which there is sensitivity in $G$ and sources would be radio-loud, but lies beyond the sensitivity limit in $t$, is labelled ``Indeterminate region''.}
\label{fig:G-t-colour_space}
\end{figure*}

Next, we determined an optimal angular separation threshold $\Delta \theta_\mathrm{t}$, for which one can safely assume that sources with $\Delta \theta < \Delta \theta_\mathrm{t}$ are physically related in Quaia and VLASS, while sources with $\Delta \theta \ge \Delta \theta_\mathrm{t}$ are plausibly just chance alignments of non-identical sources. An optimal choice of $\Delta \theta_\mathrm{t}$ should be large enough to allow for a large number of proper physical cross-matches to be identified as such (high completeness of the sample). Simultaneously, $\Delta \theta_\mathrm{t}$ should be reasonably small to prevent contamination of the cross-matched population with chance-aligned sources (high-purity sample). 

In Figure \ref{fig:ang_sep_cumulative_hist}, we present our findings in the context of the fraction of possibly false matches. In the left panel, $N_\mathrm{cross}$ is the cumulative number of cross-matches between Quaia and VLASS as a function of the coordinate separation up to the corresponding $\Delta \theta$ value. We also display $N_{\rm false}$, calculated with a method used previously by \citet{Orosz_2013}. $N_{\rm false}$ measures a cross-matched fraction between VLASS and Quaia*, where  Quaia* is a surrogate of the Quaia QSO catalog with all sources rotated randomly by the same angle in right ascension. This ensures that $N_{\rm false}$ from cross-matching Quaia* and VLASS only contains unphysical chance alignments. Moreover, $N_{\rm false}$ should approximate the chance alignments between Quaia and VLASS very well, because Quaia* has the same source density as Quaia. The curve of $N_\mathrm{corrected} = N_\mathrm{cross} - N_{\rm false}$ demonstrates a saturation in the number of physical cross-matches at $\Delta \theta \approx 1.5\arcsec$ (displayed as a grey band).

Using the above estimate of the contamination, in Figure \ref{fig:ang_sep_cumulative_hist} (right panel) we show the fraction of chance alignments as a function of $\Delta \theta$, which can be expressed as 

\begin{equation}
    f(\theta) = \frac{N_{\rm false}(\theta)}{N_{\rm corrected}(\theta)} = \frac{N_{\rm false}(\theta)}{N_{\rm cross}(\theta)-N_{\rm false}(\theta)} .
    \label{equ:frac_chance_alignments}
\end{equation}

At $\Delta \theta \approx 1.5\arcsec$, we estimated $N_{\rm false}/N_{\rm corrected} \approx 0.1\%$ contamination, i.e. excellent purity. We thus decided to apply this $\Delta \theta_\mathrm{t} < 1.5\arcsec$ threshold for our cross-matching procedure. We note that \citet{2002AJ....124.2364I} also used the same threshold for their analysis, which makes comparisons more straightforward.

Figure~\ref{fig:ang_sep_hist} shows a histogram of the Quaia--VLASS angular separations up to $\Delta \theta = 1.5\arcsec$, indicating that most cross-matches are identified with $\Delta \theta < 0.5\arcsec$ separation, with a peak at about $\Delta \theta \approx 0.07\arcsec$ (see inset of Figure \ref{fig:ang_sep_hist}). We thus conclude that the construction of our cross-matched sample guarantees high completeness. The number of the Quaia--VLASS cross-matches in different $G$ bins and in total can be looked up in Table \ref{tab:cuts_n_radio loudness}. 

\subsection{Reddening Correction}
\label{subsec:reddening-correction}
As provided by \citet{Storey_Fisher_2024}, the $G$-band magnitudes of the Quaia sources are not corrected for Galactic extinction. We thus corrected the $G$-band data following the formula
\begin{equation}
    G = G' - r \cdot \langle R_{\zeta} \rangle,
    \label{equ:mag_correction}
\end{equation}
where $G'$ is the uncorrected magnitude, $G$ is the extinction-corrected magnitude, and $r$ is the reddening in the given region of the sky as provided by \citet{1998ApJ...500..525S}. Further, $\langle R_{\zeta} \rangle \approx 2.74 $ is the extinction coefficient for the Gaia $G$-band as provided by \citet{Casagrande_2018}.
Except stated otherwise, all subsequent references in this paper to the $G$-band magnitude refer to the corrected value from Eq.~\ref{equ:mag_correction}.

\section{Results}
\label{sec:res}

Using the cross-matched Quaia--VLASS sources, we estimated the radio loudness of our sample via the definition by \citet{2002AJ....124.2364I} for the radio magnitude $t$ as

\begin{equation}
    t = -2.5 \log \left( \frac{F_\mathrm{tot}}{3631 \mathrm{Jy}} \right),
    \label{equ:t_definition}
\end{equation}
where $F_\mathrm{tot}$ is the total flux density of the source (in the $S$-band). Within the same work the radio loudness $R_m$ is expressed as the difference between the optical and radio magnitudes of the sources: $R_m = 0.4(m-t)$, where $m$ is a magnitude in the optical, $t$ is the radio magnitude at $1.4$~GHz (in that example, $R_m>1$ corresponds to radio-loud and $R_m\leq1$ to radio-quiet objects). The equation for $R_m$ turned into an inequality translates in our work to the following definitions, used throughout this paper:

\begin{equation}
    \begin{array}{ll}
    G - t > 2.5:& \text{ radio-loud sources} \\
    G - t \leq 2.5:& \text{radio-quiet sources}
    \end{array}
\label{equ:radio_loudness}
\end{equation}

\subsection{G-t color space and histogram}

The distribution of radio-loud and radio-quiet cross-matches in a $G-t$ color space is shown in Figure \ref{fig:G-t-colour_space}. Sources are colored according to bins in $G$, which correspond to optically bright, moderate, and faint bins (the number of sources within different $G$-bins is given in Table~\ref{tab:cuts_n_radio loudness}). The choice of $G$-band range values for these bins is motivated by a desire to keep these splits simple, yet offering meaningful comparisons to \citet{2002AJ....124.2364I}. Visually, the sensitivity limit in the radio lies within $16<t_\mathrm{lim}<17$. This limit can also be calculated via the conventional assumption of being equal 5 times the root mean square (RMS) noise limit \citep{2024MNRAS.528.4547A}. \cite{lacy2022vlass} provide the flux density restricted by RMS noise as $\sigma_\mathrm{fd} = 170\,\mu\mathrm{Jy}$. Inserting $F_\mathrm{tot} = 5\sigma_\mathrm{fd}$ in equation \ref{equ:t_definition} gives a magnitude limit of $t \approx 16.6$, which we plotted in Figure \ref{fig:G-t-colour_space} as a horizontal dashed line and it agrees well with the visual extent of the data points.

We treat all Quaia sources, for which we do not find a matching radio counterpart according to the methodology in Section \ref{subsec:Cross-matching}, as radio-quiet QSOs. All RLF values provided throughout this work are based on this assumption. In that way we make use of all the available optical sources, even if they do not have measured radio magnitudes. This approach provides accurate RLF values for the sources with $G \lessapprox 18.5$, because within this regime the radio loudness border at $G-t=2.5$ is above the VLASS sensitivity limit (see Figure \ref{fig:G-t-colour_space}, for all blue data points the solid line is above the dashed line). On the other hand, for sources with $G \gtrapprox 18.5$, our approach of treating radio-undetected sources as radio-quiet will lead to an underestimation of the RLF - in this regime Quaia sources could have radio-counterparts fulfilling $G-t>2.5$ that happen to have $t$ fainter than the sensitivity limit of VLASS (see the indeterminate region in Figure \ref{fig:G-t-colour_space}).

\begin{figure*}
\centering
\includegraphics[width=110mm]{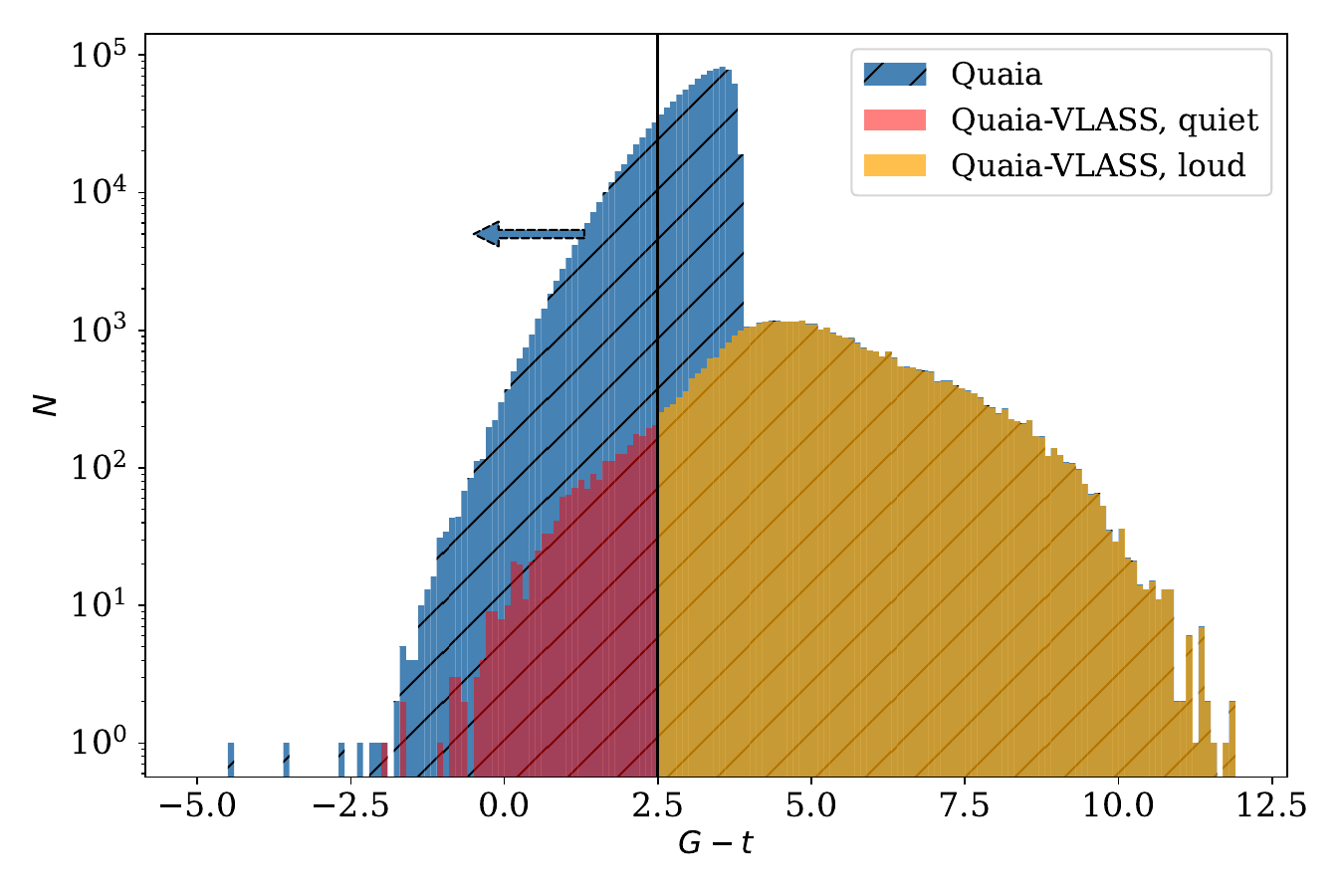}
\caption{The distribution of the radio loudness measure $G-t$ for all Quaia sources and the Quaia--VLASS cross-matches. The vertical black line is the delineation between radio-quiet ($G-t \leq 2.5$) and radio-loud ($G-t > 2.5$) sources. The blue arrow indicates that the shape of the blue distribution would shift to the left, if for these sources real measurements for $t$, instead of estimates, were present.}
\label{fig:g_t_hist}
\end{figure*}

An alternative approach to treating radio-undetected sources is presented in \cite{2024MNRAS.528.4547A}. There, the radio-undetected sources get assigned a radio magnitude equal to 5 times the local RMS value taken from the radio-survey RMS maps at the coordinates of the quasars. In the present work, this approach would result in a mean sensitivity limit in the radio of $t\approx16.6$. This method would lead to an overestimation of the RLF at faint $G$ -- expressing this in Figure \ref{fig:G-t-colour_space} would mean to populate all radio-undetected sources along the dashed line, hence the optically faintest will occupy the indeterminate region and be radio-loud. Nonetheless, this approach has the advantage of enabling a $G-t$ histogram for all sources, including the radio-undetected ones. We show this histogram of our data in Figure \ref{fig:g_t_hist}. It allows for naive check of the quasar dichotomy hypothesis -- one observes that both the radio-undetected Quaia sources (blue) and the radio-detected ones (red and orange) have smooth distributions without strong peaks indicative for subclasses in the quasar population. This is consistent with results from other recent works, reporting the transition from radio-quiet to radio-loud QSO being smooth and lacking evidence of a clear dichotomy \citep{2019A&A...622A..11G, macfarlane2021radio, 2024MNRAS.528.4547A}. Nonetheless, interpretations of Figure \ref{fig:g_t_hist} are to be done with caution, since data are limited and $t$ values for the blue histogram are only upper limits. In Section \ref{sec:disc}, we point out future surveys that may provide new insight on these patterns by using even better data sets.

\begin{figure*}
\centering
\includegraphics[width=170mm]{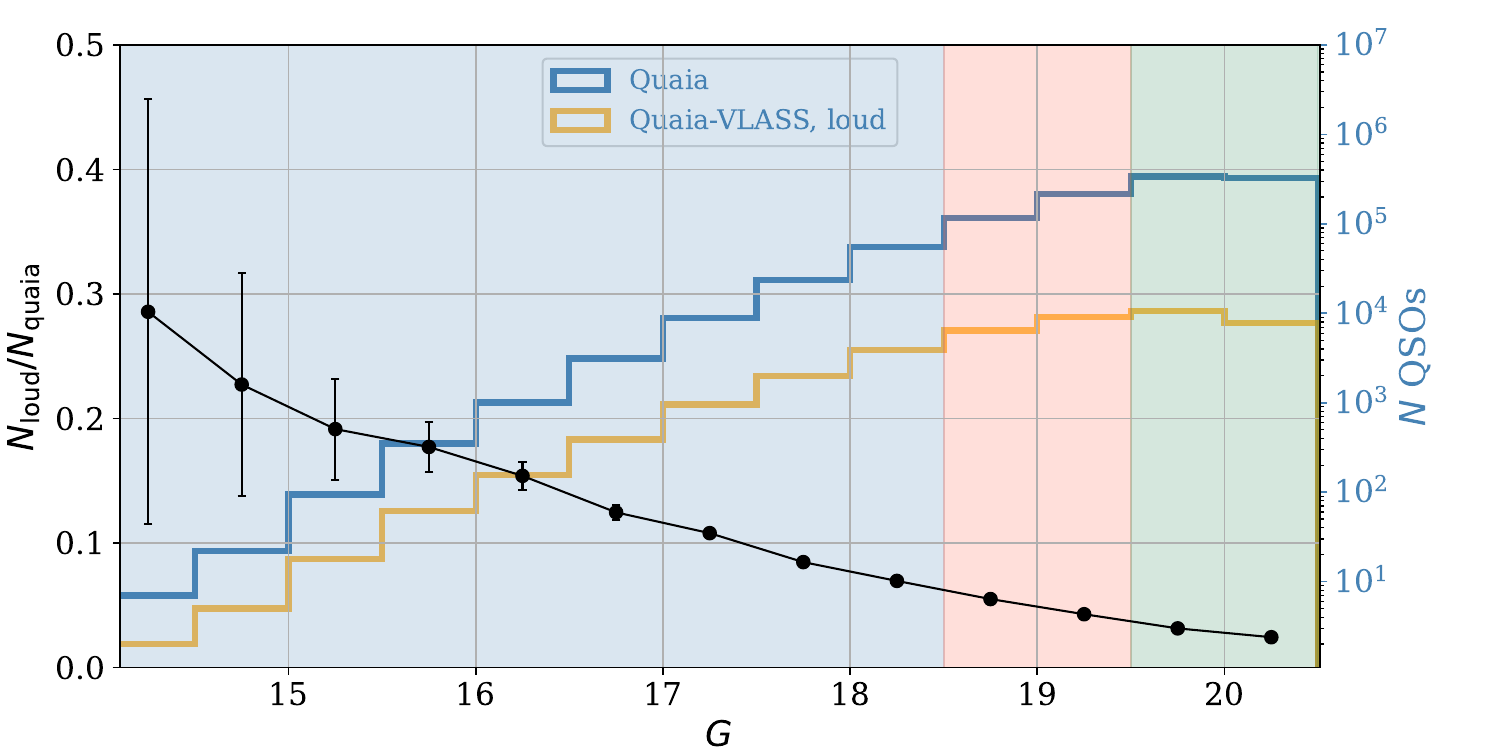}
\caption{The RLF $N_{\rm loud}/N_{\rm quaia}$ of QSOs in different bins as a function of $G$ (black curve), the number of Quaia quasars within these bins (blue histogram) and the number of radio-loud crossmatches (orange histogram). The background color-coding along $G$ is equivalent to the color-coding in Figure \ref{fig:G-t-colour_space}. The black curve shows that even though it is expected to underestimate the RLF at $G>18.5$ due to the effect of the indeterminate region (see Figure \ref{fig:G-t-colour_space}), such a trend of a sudden drop in the curve is not observed. The errors are calculated via the binomial standard error $\sqrt{f(1-f)/n_{\rm b}}$, where $f$ is the RLF in a bin and $n_{\rm b}$ the total number of QSOs in the bin.}
\label{fig:rad_loudness_vs_G}
\end{figure*}

\subsection{A High-purity Catalog of Radio-loud Quasars}

In the above Section, we kept the definitions for $t$ (Equation \ref{equ:t_definition}) and radio loudness (Equation \ref{equ:radio_loudness}) consistent with \citet{2002AJ....124.2364I} for the sake of comparison. Nonetheless, there are some inevitable technical differences, as displayed in Table~\ref{tab:comparison_previous_works}. The main contribution of this work is a significantly larger sky coverage, a larger cross-matched sample, and thus a reduction of possible biases in the analysis. 

While the actual definitions of radio loudness may differ (as well as their range of validity), we note that of the 43,650 cross-matched Quaia--VLASS quasars shown in Figure \ref{fig:G-t-colour_space} the vast majority is clearly visible in the radio (41,496 of them are radio-loud, given our definition). Therefore, an important outcome of our work is a \emph{high-purity} catalog of quasars with radio counterparts, which may be used in various other projects in the future. 

Considering the \emph{completeness} of our radio-loud catalog (besides its purity), one should also consider the sensitivity limits of the input optical and radio data sets, that may bias the radio loudness statistics for faint sources. We provide various tests of these systematic effects in the following sections, applying different binning schemes and cuts to the Quaia--VLASS sample to estimate possible variations in the RLF.

\subsection{Comparison to Previous Works}

As evident from Table \ref{tab:comparison_previous_works}, row ``$N$ cross-matches'', by using the same methodology as \cite{2002AJ....124.2364I} we discover $\approx 14$ times more optical--radio QSO cross-matches. The same table shows that our work is superior to the other compared works in terms of cross-matches due to the larger radio and optical parent catalogs we used. Many of the previous works have used SDSS QSO samples containing fainter sources than in our work. Nonetheless, all sources we observe outside the SDSS footprint are novel to the compared previous works. This means, by simply comparing the survey overlap areas in Table \ref{tab:comparison_previous_works} and making the reasonable approximation that the Quaia--VLASS sample is homogenous, our subset of optical-radio cross-matches outside the sky coverage of \cite{2012ApJ...759...30B} is $\approx 32,000$ ($\approx 30,000$ of them radio-loud). Likewise, our subset of cross-matches outside of the \cite{2024MNRAS.528.4547A} sky area is $\approx 35,000$ ($\approx 33,000$ of them radio-loud). By covering a significantly larger sky area, our catalog is superior to the compared samples, for users desiring the closest possible representation of the full sky at present. Also, in contrast to some of the previous works, our code and final data products are freely available to the public, see Section~\ref{sec:final_cat_and_data_avail}. Our catalog is also superior for cosmological applications, since the Gaia scanning pattern is more homogenous than selective SDSS QSO observations and the Quaia selection function can help further mitigate selection biases. On the other hand, the SDSS spectroscopic observations are superior to the photometric ones by Gaia, so redshift and classification accuracy are better in the previous studies in Table \ref{tab:comparison_previous_works}.

\begin{table}
\raggedright  \caption{Comparison of Quasar radio loudness Studies}
\hspace*{-1cm}  \resizebox{19.cm}{!}{\begin{tabular}{l c c c c c}
\hline \hline
\textbf{Parameter} & \textbf{\cite{2002AJ....124.2364I}} & \textbf{\cite{kimball2008unified}} & \textbf{{\cite{balokovic2012disclosing}}}& \textbf{\cite{2024MNRAS.528.4547A}} & \textbf{this work} \\
\hline
Cross-match area [deg$^2$] & 1,230 & 2,894 & 7,600 & 5,634 & 29,230 \\
$N$ optical sources$^a$ & 25.3M (all object types) & 35,450 & 98,544 &189,680 & 1.085M \\
$N$ radio sources$^b$ & 107,654 & 140,000 & -- & 4.4M (in full LoTSS area) & 1.874M \\
$N$ cross-matches & 3,066 & 2,927 & 8307 & 33,968 & 43,650 \\
$N$ cross-matches, loud & -- & -- & -- & 3,697$^c$ & 41,496\\
Optical survey (band) & SDSS ($i$, $r$) & SDSS DR5 ($i$, $g$, $r$) & SDSS DR7 ($i$) & SDSS DR14 ($i$) & Gaia ($G$) \\
Optical limit & $i<21$ & $i\lesssim21$ & $i\lesssim21.2$ &$i\lesssim21$ & $G<20.5$ \\
Radio survey (frequency) & FIRST (1.4 GHz) & NVSS, FIRST (both 1.4 GHz) & FIRST (1.4 GHz) & LoTSS DR2 (144 MHz) & VLASS2 QL (3 GHz) \\
Radio limit$^d$ & 1 mJy ($t\approx16.4$) & 1 mJy ($t\approx16.4$) & 1 mJy ($t\approx16.4$) & 0.415 mJy ($t \approx 17.35$) & 0.85 mJy ($t\approx 16.6$) \\
\hline
\end{tabular}
}
\tablecomments{\\
$^a$ Number of quasars (except different object type noted in parentheses) from the optical survey located inside the cross-match area.\\
$^b$ Number of sources from the radio survey located inside the cross-match area (except different area is noted in parentheses).\\
$^c$ \cite{2024MNRAS.528.4547A} use a different definition for radio loudness.\\
$^d$ The magnitude limit $t$ is calculated by inserting the flux density limit in Equation \ref{equ:t_definition}.\\}
\label{tab:comparison_previous_works}
\end{table}

\subsection{Radio-loud Fraction vs. \texorpdfstring{$G$}{G} Magnitude}

When estimating the RLF ($N_{\rm loud}/N_{\rm total}$) of Quaia--VLASS sources, it is reasonable to restrict the sample to a range where the sensitivity of the instruments is reliable, i.e. where the measured $N_{\rm loud}$ and $N_{\rm total}$ source counts are complete within a chosen magnitude limit. 

\citet{2002AJ....124.2364I} applied such a restriction in the form of a cut in the optical $i$-band, providing a fiducial value for the fraction of radio-loud QSOs with $i<18.5$. This particular cut follows from their radio loudness formula ($G-t>2.5$) and the sensitivity limit they work with in the radio ($t<16.0$, that is slightly more conservative than the nominal $t<16.5$ where they still find radio sources). 

\begin{figure*}
\centering
\includegraphics[width=165mm]{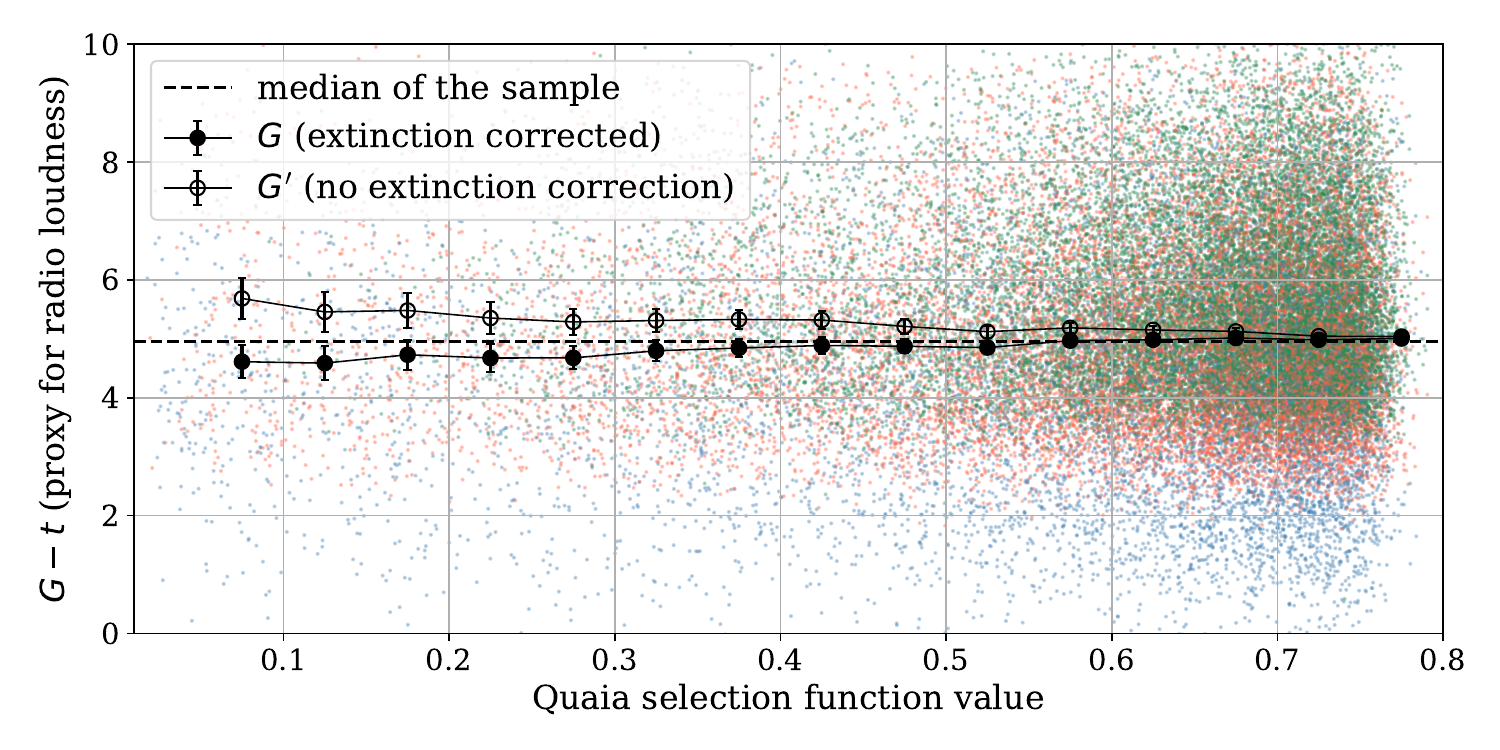}\\
\includegraphics[width=165mm]{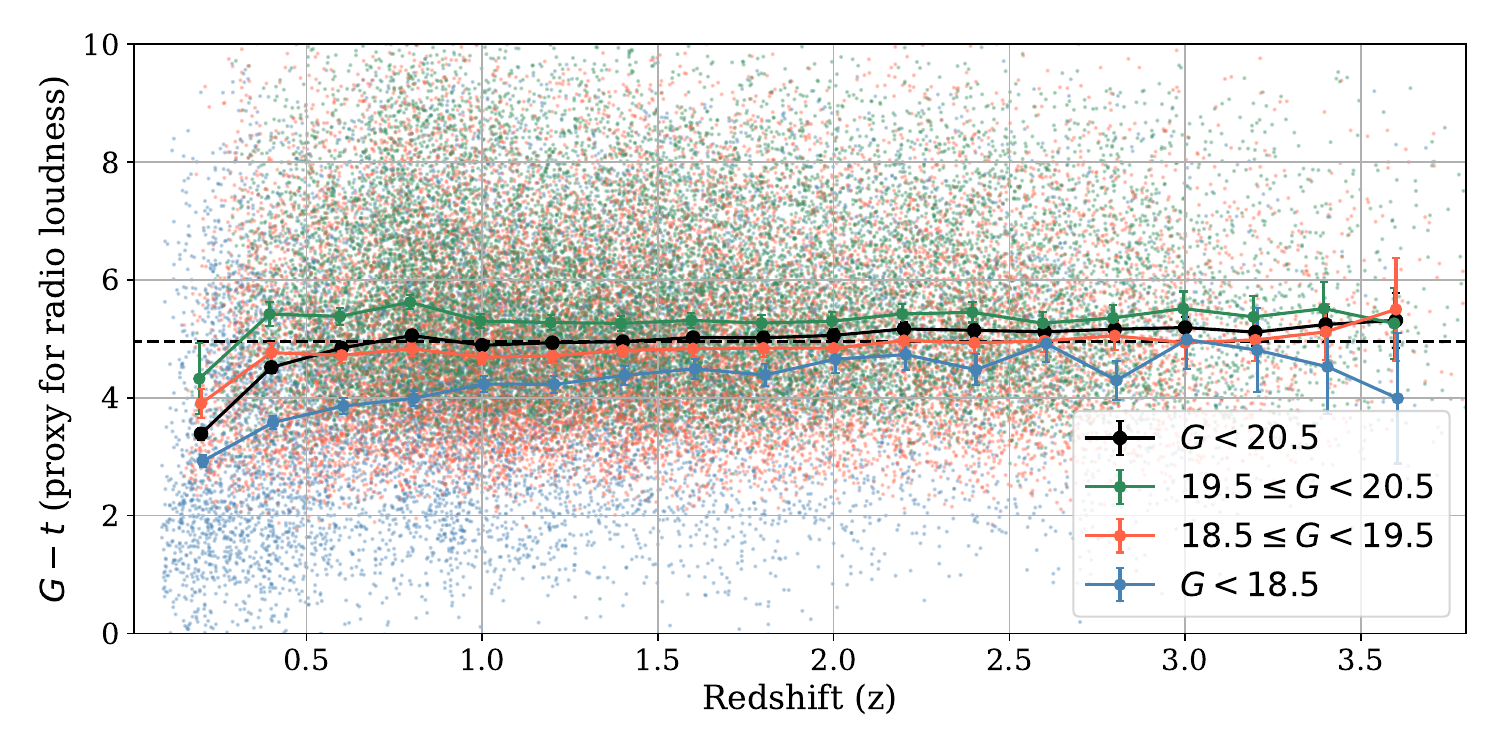}
\caption{The $G-t$ magnitude difference, i.e. our proxy for radio loudness, and its median in various bins. All error bars are equal to $\mathrm{median}(G-t)/\sqrt{n_{\rm c}}$ with $n_{\rm c}$ being the number of cross-matched QSOs in this bin and $\mathrm{median}(G-t)$ being the median within the same bin. \textit{Top:} we bin our catalog on the Quaia selection function (completeness) values, and also illustrate the role of the Galactic extinction correction. \textit{Bottom:} we bin our sample on redshift, and also on $G$ magnitudes (see colors), and identify weak trends at the lowest redshifts $(z<0.5)$.}
\label{fig:sel_func_vs_G-t}
\end{figure*}

In our work, we used the same Eq.~\ref{equ:radio_loudness} for defining radio loudness, and the VLASS has a similar sensitivity limit in $t$ (see Figure \ref{fig:G-t-colour_space}). While this condition motivates a cut in the Quaia data at $G<18.5$, we tested the whole $G<20.5$ magnitude range to measure $N_{\rm loud}/N_{\rm total}$ for a comprehensive analysis.

Figure~\ref{fig:G-t-colour_space} depicts the origin of this unreliability. For fainter Quaia magnitudes, the radio magnitudes may fall outside of the sensitivity limit $t\approx16.0$, which suggests lower and lower completeness when one intends to estimate the RLF on the faint end (while the purity of the sample is expected to remain high, as we argued above).

In Figure \ref{fig:rad_loudness_vs_G}, we show the RLF of QSOs as a function of $G$ (denoted by the black line), where we make the following observations:

\begin{itemize}
    \item The RLF of sources drops linearly for fainter optical sources (which are naturally more abundant in the sample).
    \item the RLF is not constant even in the range of the brightest sources (e.g. $G < 17$), where high completeness is expected in both $N_{\rm loud}$ and $N_{\rm total}$.
    \item Constraining the measurement of the RLF to $G < 18.5$ would ignore the bulk of the Quaia QSOs (see the blue histogram in \ref{fig:rad_loudness_vs_G}) and the majority of the cross-matched sources (see Table~\ref{tab:cuts_n_radio loudness}).
\end{itemize}

For the above three reasons, we decided to express the RLF of the Quaia--VLASS sources as a function of $G$ up to $G=20.5$, and let future users of our catalog decide how they wish to balance purity and completeness.

\begin{figure*}
\centering
\includegraphics[width=155mm]{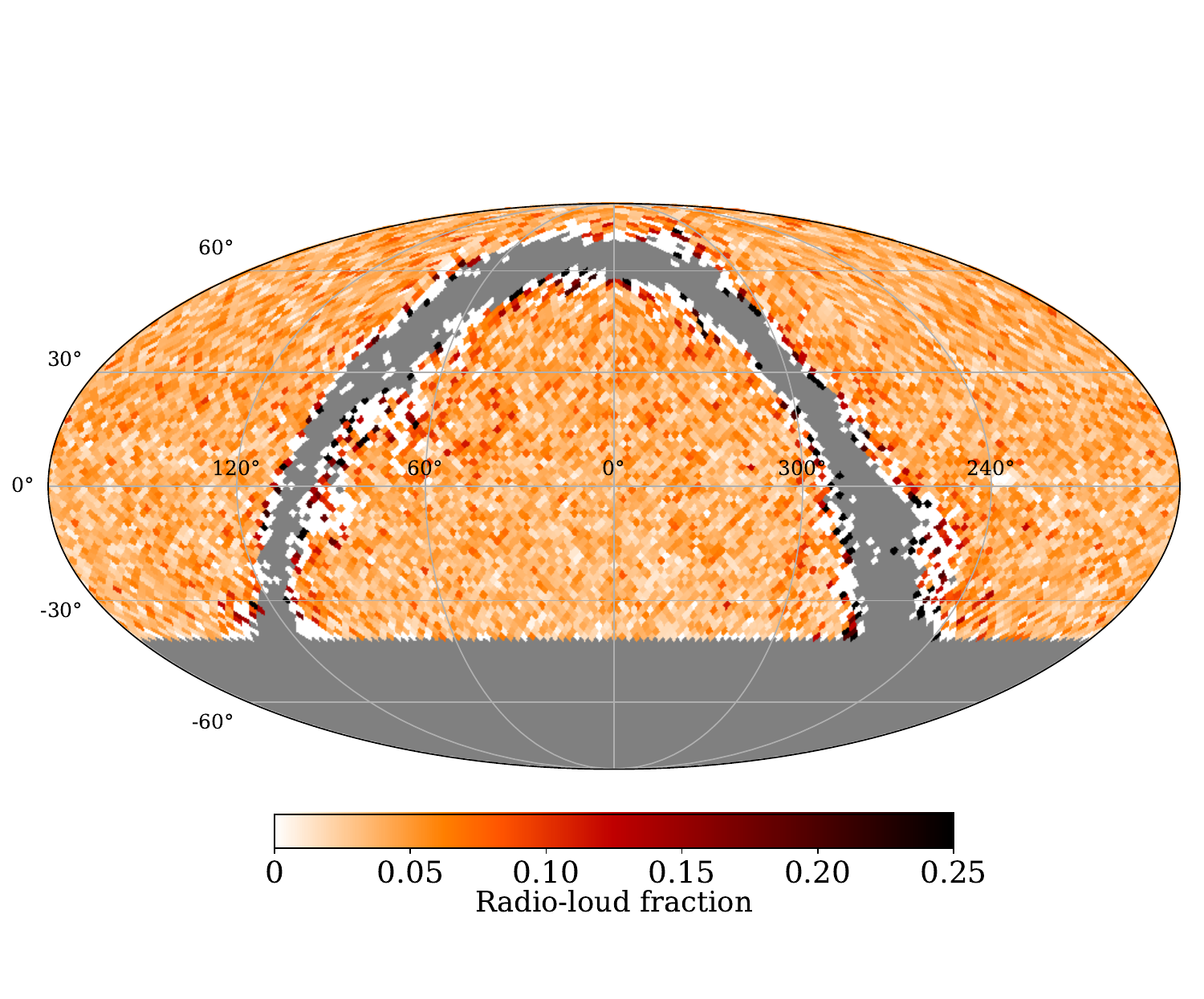}
\caption{Quasar RLF in different parts of the sky. We find significant fluctuations near the Milky Way, i.e. where the Quaia catalog is knowingly less complete (although on average there are no clear biases near the Galactic plane either, see top panel of Figure \ref{fig:sel_func_vs_G-t}). In the cleanest areas of the sky, we do not identify any relevant anisotropy in the RLF.}
\label{fig:rad_loudness_sky}
\end{figure*}

Nevertheless, if we apply the measure for the RLF by \citet{2002AJ....124.2364I}, then we arrive at a value of $N_{\rm loud}/N_{\rm total} \approx 8.04\% \pm 0.09\%$ RLF for $G<18.5$, while they estimate $N_{\rm loud}/N_{\rm total} \approx 8\% \pm 1\%$ for a similarly bright, conservatively selected subset of their data ($i<18.5$). Again, we note the differences between the actual bands used by \citet{2002AJ....124.2364I} and our analysis, but we conclude that consistent RLF values are observed for the brightest QSOs.

\subsection{Radio Loudness vs. QSO Completeness, Redshift, and Sky Coverage}
\label{sec:Radio_Loudness_vs_QSO_Completeness_and_Redshift}

In Figure \ref{fig:sel_func_vs_G-t}, we show multiple tests that we performed by binning and modifying our data. In the context of the Quaia selection function (a measure for the completeness of QSO detections per pixel), we show on the top panel the role of the $G$-band extinction correction to $G-t$ (our proxy for the radio loudness), and the overall spread of the Quaia--VLASS catalog as a function of this systematic parameter of the input data. 

The color of the data points corresponds to their $G$-band magnitudes (as shown in Figure \ref{fig:G-t-colour_space}). Empty circles mark the median value of $G'-t$ within a given bin of the selection function values, considering the raw $G'$ magnitudes (not corrected for extinction). In contrast, full black circles show the binned median $G-t$ values after extinction correction. 

From the comparison to the median $G-t$ value of the whole Quaia--VLASS sample (horizontal line in top of Figure \ref{fig:sel_func_vs_G-t}), we found that the extinction correction is effective in reducing the deviations between $G-t$ values measured at low vs. high selection function values. The top panel of Figure \ref{fig:sel_func_vs_G-t} demonstrates that most filled circles fall within $1\sigma$ of the global median $G-t$ value for all cross-matches. This result further justifies our use of the extinction-corrected $G$-band values in subsequent analyses, and it suggests that Galactic dust extinction will not significantly affect our radio loudness analysis. 

To perform a further test of systematics, we looked for possible changes in radio loudness as a function of QSO redshift. As shown in Figure \ref{fig:sel_func_vs_G-t} (bottom panel), we observed a moderately significant bias in the median $G-t$ distribution (i.e. in the proxy for radio loudness) at the lowest redshifts ($z<0.5$). We estimated this statistic for the three bins in G magnitude that we considered above, and we found almost identical trends as for the $G<20.5$ case.

The source of this biased $G-t$ distribution at the low-redshift part of the Quaia QSO catalog remains unclear. One possibility is that the star--galaxy--QSO separation in the Gaia--WISE color space is less efficient, or the photometric redshifts of the quasars might also be biased at the lowest redshifts where the training set based on Sloan Digital Sky Survey (SDSS) data is less complete. For a detailed explanation, further measurements and statistical probes would be required, which are beyond the scope of this paper.

\begin{figure*}
\centering
\includegraphics[width=165mm]{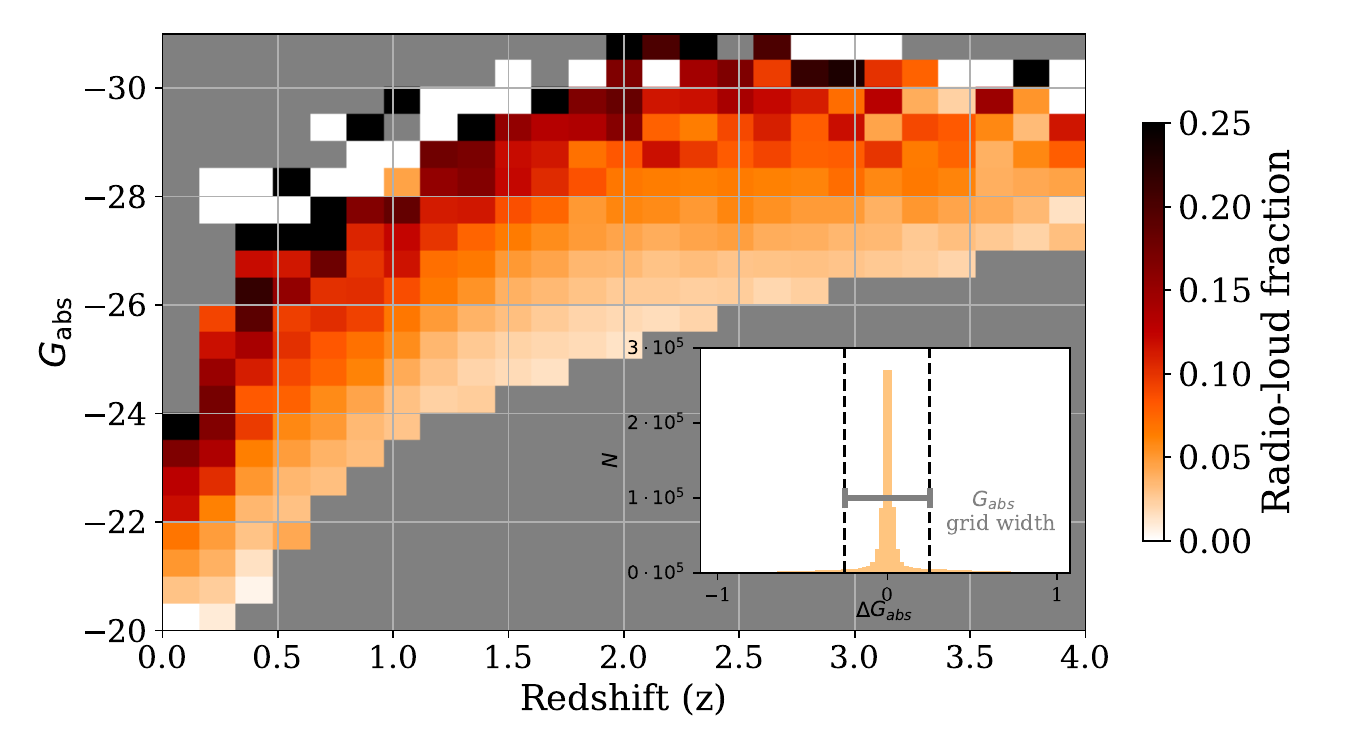}\\
\caption{RLF of Quaia sources color-coded as a function of the absolute magnitude $G_{\rm abs}$ and the redshift $z$. We observed significant noise due to small-number statistics at the bright-end pixels, leading to strong fluctuations in the RLF. The inset shows the spread of $\Delta G_\mathrm{abs}$ being comparatively small to the grid-width of $G_\mathrm{abs}$ in the large figure.}
\label{fig:r_loudnes_vs_gabs_vs_z}
\end{figure*}

Looking for possible large-scale spatial patterns on the sky, we created a pixelized map of the radio-loud QSO fraction within each pixel. In Figure \ref{fig:rad_loudness_sky}, a rather homogeneous distribution is shown across the full observed area, except close to the Galactic plane. 

We note that the fluctuations in radio loudness are naturally stronger at low Galactic latitudes, due to sparsity of the QSOs (see left side of Figure \ref{fig:sky_map_quaia_vlass}). For a conservative analysis, future users of our Quaia--VLASS catalog may apply specific cuts in the Quaia selection function map, offering convenient ways of limiting the analyses to the cleanest and most complete parts of the sky.

\subsection{Radio-loud Fraction vs. Absolute Magnitude}

Additionally, we tested the possible dependence of the RLF on the \emph{absolute} magnitudes of the quasars. Following \citet{2002AJ....124.2364I}, we used the absolute magnitude $G_{\rm abs}$, which we calculated as
\begin{equation}
    G_{\rm abs} = G + 5 - 5 \cdot \log\left(d_{\rm lm}(z)\right),
    \label{equ:G_abs_calc}
\end{equation}
where $d_{\rm lm}(z)$ is the luminosity distance of the object. 

We highlight here that we consider this as an exploratory test, because the Quaia photometric redshifts have a non-negligible uncertainty compared to more ideal spectroscopic redshifts of QSOs. These errors also propagate to our estimation of the luminosity distances, and therefore to the $G_{\rm abs}$ values.

Given the above caveats, our main findings regarding the RLF patterns in $G_{\rm abs}$ and redshift bins are illustrated in Figure \ref{fig:r_loudnes_vs_gabs_vs_z}. Each pixel displays the RLF of Quaia sources in that bin,  using a color-coding scheme to easily identify any possible trend in the data along either axis. 

The figure allows a separate analysis of constant $z$ values in rows against varying $G_{\rm abs}$ values in columns, or vice versa. We found that for a fixed redshift value, the RLF increases towards brighter $G_{\rm abs}$ bins. Similarly, for a fixed $G_{\rm abs}$, the binned RLF values show a decreasing trend with increasing redshift. Applying a partial Spearman correlation to calculate the significance of the observed trends yields a correlation of the RLF with redshift (controlling for $G_{\rm abs}$) with values $r \approx -0.80$ and  $p \approx 2.07 \times 10^{-52}$ and a correlation of the RLF with $G_{\rm abs}$ (controlling for redshift) with values $r \approx -0.83$ and  $p \approx 1.48 \times 10^{-58}$. Hence, we can rule out that the trends are caused due to chance noise.

Below we discuss how much redshift errors influence the trends in Figure \ref{fig:r_loudnes_vs_gabs_vs_z}. The redshift error stated by \cite{Storey_Fisher_2024} for the full Quaia catalog is $\mathrm{mean(|\Delta z/(1+z)|)\approx0.01}$, significantly smaller than the grid size along $z$ in Figure \ref{fig:r_loudnes_vs_gabs_vs_z}. \cite{Storey_Fisher_2024} assign the final redshift $z$ of a source, based on conditions, between $z=z_{kNN}$ (the redshift they estimate from a kNN method) and $z=z_{\rm Gaia}$ (the raw redshift for Quaia coming from \cite{2023A&A...674A..31D}). We follow the same conditions to pick for each source its corresponding redshift error standard deviation $\sigma_z$ from the Quaia catalog: $\sigma_z=$\textit{redshift\_spz\_err}, if $z=z_{kNN}$, and $\sigma_z=$\textit{redshift\_qsoc\_upper}, if $z=z_{\rm Gaia}$. Using these standard deviations, we assign to each source in Quaia probabilistically a new, erroneous redshift $z_*$ and corresponding $G_{\rm abs}(z_*)$, and then calculate $\Delta G_{\rm abs} = G_{\rm abs}-G_{\rm abs}(z_*)$, the fluctuation in absolute magnitude due to redshift uncertainties. $\Delta G_{\rm abs}$ is plotted as inset in Figure \ref{fig:r_loudnes_vs_gabs_vs_z} and one can see that its spread is small compared to the $G_{\rm abs}$ grid size of 0.5 magnitudes in the same figure, and so are $\mathrm{median}(|\Delta G_{\rm abs}|)\approx 0.02$ and $\mathrm{mean}({|\Delta G_{\rm abs}|}) \approx 0.17$. Only $\approx17.3\%$ of the sources have a $\Delta G_{\rm abs}$ being outside of the 0.5 magnitude grid size width drawn with vertical lines in the inset plot. Hence, we conclude that redshift errors and their resulting $G_{\rm abs}$ errors are too small, compared to the grid size in both dimensions of Figure \ref{fig:r_loudnes_vs_gabs_vs_z}, to significantly influence trends observable in the figure.

Previous works report controversial results for the relations of RLF with luminosity and redshift. Some authors find no (or weak) correlation between RLF and both luminosity and redshift \citep{stern2000radio, 2002AJ....124.2364I, liu2021constraining, banados2015constraining, keller2024radio}. The latter two studies discuss redshift only. According to other researchers, RLF increases with decreasing redshift and increasing optical/UV luminosity \citep{jiang2007radio, kratzer2015mean, rusinek2021dependence, lah2023ultraluminous}. In general, RLF can be influenced by the choice of: radio and optical bands; radio and optical flux limits; the exact value of the radio loudness threshold parameter; AGN variability, which affects mostly objects with radio loudness close to the demarcation value. The observed dependence of RLF on redshift could be explained e.g. in terms of the Mixture Evolution Scenario of the AGN Radio Luminosity Function \citep{yuan2017mixture}.

We note that, besides the known redshift errors, our $G_{\rm abs}-z$ plot in Figure~\ref{fig:r_loudnes_vs_gabs_vs_z} displays significant noise due to the sparsity of objects in pixels with the most negative $G_{\rm abs}$ values, leading to large fluctuations in RLFs. Hence, we suggest that QSOs falling in these pixels should be flagged or ignored in any subsequent analyses of redshift- and magnitude-dependence of RLF that use our Quaia--VLASS catalog.

\begin{figure*}
\centering
\includegraphics[width=180mm]{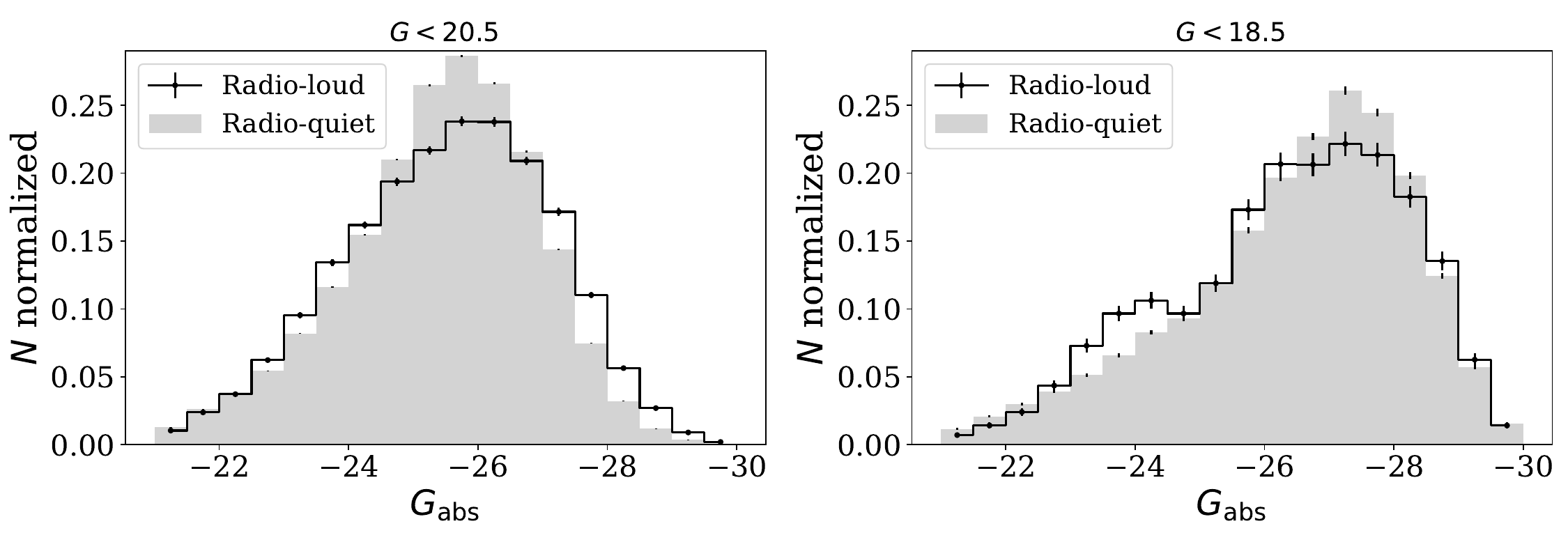}
\caption{Normalized histograms of the radio-loud and radio-quiet populations as a function of $G_{\rm abs}$. On the left, we show the full sample of cross-matches, while on the right a $G<18.5$ cut is applied. The error bars in the figure are Poisson errors equal to $n_{\rm norm}/\sqrt{n_0}$ with $n_{\rm norm}$ being the displayed normalized counts within the bin and $n_0$ the non-normalized counts in the same bin.}
\label{fig:G_abs_hist}
\end{figure*}

\subsection{Radio-loud vs. Radio-quiet Source Characteristics}

In this Section, we mostly followed \citet{2002AJ....124.2364I} and explored if radio-loud and radio-quiet QSOs are statistically consistent with each other, including differently binned subsets. To this end, they compared the normalized histograms of radio-loud and radio-quiet sources as a function of the $i$-band absolute magnitude (similar to our Figure \ref{fig:G_abs_hist}). Above all, they identified an excess of radio-quiet QSOs in the faintest bins, followed by a more abundant radio-loud population, and then again more radio-quiet sources at the peak of the distribution. 

\citet{2002AJ....124.2364I} attributed these features to noise fluctuations and claimed that differences might only be caused by a few dozen sources. In our analysis, we created similar histograms (see the right panel in Figure \ref{fig:G_abs_hist}) using $G_{\rm abs}$ (limited to objects with $G<18.5$), as opposed to $i_{\rm abs}$ and an $i<18.5$ cut in their work. Further, we analyzed more sources with 7,431 radio-loud QSOs at $G<18.5$, compared to their 280 radio-loud sources at $i<18.5$. Consequently, our larger data set is expected to help clarify the nature of the excess radio-loud populations in some $G_{\rm abs}$ bins, that were identified as random patterns in their work.

Considering the brighter $G<18.5$ subset of our Quaia--VLASS quasars, we found that an excess of radio-loud sources does persist at $G_{\rm abs}\approx-24$, and given the statistical errors in our sample, this pattern certainly reaches several standard deviations in three neighbouring bins (see the right panel in Figure \ref{fig:G_abs_hist}).

On the left side of Figure \ref{fig:G_abs_hist}, we present an analogous histogram for the full Quaia--VLASS sample ($G<20.5$) for a more inclusive analysis. We found that the excess of radio-loud sources persists even in the full sample, albeit in a less pronounced manner. In this larger catalog, we also noted a similar radio-loud excess for the brightest sources at $-30<G_{\rm abs}<-27$. Identifying its origin and exact significance is beyond the scope of this paper, as they might be attributed to yet-unknown systematic effects in the data, or they might be sourced by the Quaia photometric redshift errors.

\section{Summary and conclusions}
\label{sec:disc}

Driven by the availability of modern survey data sets to improve our overall understanding of the astrophysical and cosmological properties of QSOs, we cross-matched the optically selected Quaia (Gaia+WISE) catalog of quasars with the VLASS radio sources. In our methodology, we closely followed \cite{2002AJ....124.2364I} who also analysed the properties of radio-loud QSOs in a similar context, but using significantly smaller data sets. Our main results are the following:

\begin{enumerate}
    \item We created a Quaia--VLASS catalog of 43,650 QSOs at declinations $\delta > -40\degr$ with matched radio counterparts, and we make this new catalog publicly available for further analyses.
    \item This new Quaia--VLASS catalogue provides $\approx$ 14 times more QSOs compared to \cite{2002AJ....124.2364I} across $\approx 24$ times larger sky area. Similarly, due to larger sky coverage, we provide $\approx 32,000$ optical-radio QSO cross-matches outside of the sky area of \cite{balokovic2012disclosing}, and $\approx 35,000$ cross-matches outside of \cite{2024MNRAS.528.4547A}.
    \item We measured the radio loudness as a function of the $G$ apparent magnitude, and found consistent results with \cite{2002AJ....124.2364I} considering a similar magnitude limit ($G<18.5$).
    \item We identified weak trends of radio loudness changes as a function of the Quaia selection function pointing to possible systematic effects as sources.
    \item We followed up on the findings by \cite{2002AJ....124.2364I} in the context of $G_{\rm abs}$ absolute magnitude statistics, and confirmed with higher statistical significance the previously seen excess of radio-loud quasars at about $G_{\rm abs}\approx-24$, for the brightest optical sources ($G<18.5$).
    \item We find that RLF increases with luminosity and decreases with redshift as opposed to previous studies in the literature, but this could be affected by a loss of completeness with increasing redshift and decreasing luminosity.
    \item We analyse the $G-t$ distribution of sources and find, given the constraints of our data, no evidence for a radio-loudness-dichotomy
\end{enumerate}

In summary, here we provided a detailed presentation of the Quaia--VLASS catalog, and performed various statistical analyses in different subsets of the data to spot possible discrepancies or unexpected trends. While our findings already provided new insights on the radio properties of quasars, the upcoming release of the VLASS Epoch 3 catalog and a stacked catalog of the first three epochs (currently under construction) will facilitate an even more advanced analysis. With sources much fainter in the radio, we will further explore the dichotomy between radio-loud and radio-quiet sources, as well as possible relations between the positions of the QSOs in the large-scale cosmic web around them, among other potential applications.

\section{Catalog of Quaia--VLASS cross-matches and data availability}
\label{sec:final_cat_and_data_avail}
The catalog of Quaia--VLASS cross-matches obtained in this work is depicted in Table~\ref{tab:cross_match_cat}, which shows column names, units, column descriptions, origin of the column and one example entry. More information about the sources can be obtained by joining them with the original catalogs via the associated identiﬁer (Gaia, unWISE or VLASS). The cross-matches catalog is publicly available at the webpage of the online publication. It is also available with open-source code for replicating the catalog, this paper's results, and figures at \href{https://doi.org/10.5281/zenodo.16035690}{10.5281/zenodo.16035690}.

Further, the Quaia quasar catalog, its mask and selection function are made by \cite{Storey_Fisher_2024} publicly available \citep{zenodo.10403370}\footnote{\url{https://doi.org/10.5281/zenodo.10403370}}, as well as the VLASS radio source catalog\footnote{\url{https://cirada.ca/vlasscatalogueql0}}.

\begin{table}[htbp]
    \centering
    \caption{Quaia--VLASS cross-matches catalog}
    \label{tab:cross_match_cat}
    \begin{tabular}{lccllc}
        \hline \hline
        Column Name & Symbol & Unit & Description & From & Example Entry Value \\ \hline
        
        \texttt{source\_id} & ... & ... & Gaia DR3 source identifier & Quaia & 2846730608989264640 \\
        \texttt{unwise\_objid} & ... & ... & unWISE DR1 source identiﬁer & Quaia & 0000p212o0014594 \\
        \texttt{redshift\_quaia} & $z$ & ... & spectrophotometric redshift estimate & Quaia & 0.9070 \\
        \texttt{redshift\_quaia\_err} & ... & ... & 1$\sigma$ uncertainty on spectrophotometric redshift estimate & Quaia & 0.0821 \\
        \texttt{ra} & ... & deg & barycentric R.A. of the source in ICRS at 2016.0 & Quaia & 0.006143428 \\ 
        \texttt{dec} & ... & deg & barycentric declination of the source in ICRS at 2016.0 & Quaia & 21.276711143 \\ 
        \texttt{phot\_g\_mean\_mag} & $G'$ & mag & Gaia $G$-band mean magnitude (not ext. corrected) & Quaia & 20.197878 \\
        \texttt{phot\_bp\_mean\_mag} & ... & mag & Gaia integrated BP mean magnitude & Quaia & 20.221630 \\
        \texttt{phot\_rp\_mean\_mag} & ... & mag & Gaia integrated RP mean magnitude & Quaia & 19.663361 \\
        \texttt{mag\_w1\_vg} & ... & mag & unWISE W1 magnitude & Quaia & 15.583867 \\
        \texttt{mag\_w2\_vg} & ... & mag & unWISE W2 magnitude & Quaia & 14.950078 \\

        \texttt{Component\_name} & ... & ... & unique VLASS name & VLASS & VLASS2QLCIR \\
        \texttt{} &  &  &  &  & J000001.47+211636.0 \\ \texttt{RA} & ... & deg & right ascension in VLASS & VLASS & 0.00614736 \\
        \texttt{DEC} & ... & deg & declination in VLASS & VLASS & 21.276691 \\
        \texttt{Total\_flux} & $F_\mathrm{tot}$ & mJy & integrated radio flux density & VLASS & 1.846 \\

        \texttt{reddening} & $r$ & ... & sky area's reddening (Eq.~\ref{equ:mag_correction}), from \citet{1998ApJ...500..525S} & this work & 0.052644 \\
        \texttt{t} & $t$ & mag & radio magnitude (Eq.~\ref{equ:t_definition}) & this work & 15.734417 \\
        \texttt{g} & $G$ & mag & extinction corrected $G$-band magnitude (Eq.~\ref{equ:mag_correction}) & this work & 20.053634 \\
        \texttt{g\_abs} & $G_\mathrm{abs}$ & mag & absolute extinction corrected $G$-band magnitude (Eq.~\ref{equ:G_abs_calc}) & this work & $-23.845170$ \\
        \texttt{sep} & $\Delta \theta$ & arcsec & angular separation between VLASS and Quaia coordinates & this work & 0.07372 \\
        \texttt{r\_loudness} & ... & ... & radio-loud (1) or radio-quiet (0) (Eq.~\ref{equ:radio_loudness}) & this work & 1 \\
         
        \hline
    \end{tabular}
\tablecomments{For an example entry, we show the ﬁrst catalog row.\\
This table is freely available in its entirety in machine-readable form.}
\end{table}

\section*{Acknowledgments}

The authors thank the anonymous referee for their helpful and constructive suggestions, which improved the clarity and quality of this manuscript. Further thanks are due to Kate Storey-Fisher and the Quaia team for their feedback and assistance with the input quasar catalog. The Large-Scale Structure (LSS) research group at Konkoly Observatory has been supported by a \emph{Lend\"ulet} excellence grant by the Hungarian Academy of Sciences (MTA). This project has received funding from the European Union’s Horizon Europe research and innovation programme under the Marie Skłodowska-Curie grant agreement number 101130774. Funding for this project was also available in part through the Hungarian National Research, Development and Innovation Office (NKFIH, grants OTKA NN147550 and K134213). S.F. was supported by the NKFIH excellence grant TKP2021-NKTA-64. L.S.-M. was partially supported by the Bulgarian Ministry of Education and Science under Agreement D01-326/04.12.2023.

\section*{ORCID IDs}
\noindent
Nestor Arsenov: \href{https://orcid.org/0009-0009-8592-5400}{0009-0009-8592-5400} \\
S\'andor Frey: \href{https://orcid.org/0000-0003-3079-1889}{0000-0003-3079-1889} \\
Andr\'as Kov\'acs: \href{https://orcid.org/0000-0002-5825-579X}{0000-0002-5825-579X} \\
Lyuba Slavcheva-Mihova: \href{https://orcid.org/0000-0002-1582-4913}{0000-0002-1582-4913}

\bibliographystyle{aasjournal}
\bibliography{ms_20250718}

\end{document}